\documentclass[useAMS,usenatbib,usegraphicx]{mn2e}

\usepackage{paper_def}
\usepackage{txfonts}
\title[Evidence of strong quasar feedback in the early Universe]{Evidence of strong quasar
feedback in the early Universe}
\author[Maiolino et al.]{R.~Maiolino,$^{1}$
S.~Gallerani,$^{2}$ R.~Neri,$^{3}$ C.~Cicone,$^{1}$ A.~Ferrara,$^{2}$
R.~Genzel,$^{4}$
D.~Lutz,$^{4}$
\newauthor
E.~Sturm,$^{4}$ L.~J.~Tacconi,$^{4}$
F.~Walter,$^{5}$ C.~Feruglio,$^{3}$ F.~Fiore,$^{6}$ E.~Piconcelli $^{7}$
\\
$^{1}$Cavendish Laboratory, University of Cambridge, 19 J. J. Thomson Ave., Cambridge CB3 0HE, UK\\
$^{2}$Scuola Normale Superiore, Piazza dei Cavalieri 7, 56126 Pisa, Italy\\
$^{3}$Institute de Radioastronomie Millimetrique (IRAM), 300 Rue de la Piscine,
38406, St. Martin d'Heres, Grenoble, France\\
$^{4}$Max-Planck-Institut f\"{u}r Extraterrestrische Physik (MPE),
	Giessenbachstr. 1, 85748 Garching, Germany\\
$^{5}$Max-Planck-Institut für Astronomie, Königstuhl 17, 69177 Heidelberg,
	Germany\\
$^{6}$Osservatorio Astronomico di Roma, INAF, via di Frascati 33, 00040 Monteporzio Catone, Italy\\
$^{7}$ESA-ESAC, Camino bajo del Castillo, Villanueva de la Canada, E-28692 Madrid, Spain 
}
\begin{document}

\date{Accepted . Received}

\pagerange{\pageref{firstpage}--\pageref{lastpage}} \pubyear{2002}

\maketitle

\label{firstpage}

\begin{abstract}
Most theoretical models invoke quasar driven outflows to quench star formation
in massive galaxies, and this feedback mechanism is required to account for the
population of old and passive galaxies observed in the local universe. The
discovery of massive, old and passive galaxies at z$\sim$2,
implies that such quasar
feedback onto the host galaxy must have been at work very early on, close to
the reionization epoch.
We have observed the [CII]158$\mu$m
transition in SDSSJ114816.64+525150.3 that, at z=6.4189, is one of the most
distant quasars known. We detect broad wings of the line tracing a
quasar-driven massive outflow. This is the most distant massive outflow
ever detected and is likely tracing the long sought quasar feedback, already at
work in the early Universe.
The outflow is marginally resolved on scales of $\sim$16~kpc, implying that
the outflow can really affect the whole galaxy, as required by quasar
feedback models.
The inferred outflow rate, $\rm \dot{M}> 3500~M_{\odot}~yr^{-1}$, is the
highest ever found. At this rate the outflow 
can clean the gas in
the host galaxy, and therefore quench star formation, in a few million
years.
\end{abstract}


\section{Introduction}

The old stellar population and low gas content characterizing local massive
galaxies, as well as their steeply declining number at high masses, have been
difficult to explain in most theoretical scenarios of galaxy formation. Indeed,
in the absence of ``negative feedback'' on star formation (i.e. a mechanism to decrease
star formation), theoretical models
expect a much larger number of massive galaxies than observed and characterized
by young stellar populations \citep{Menci2006,Bower2006}.
To reconcile theory with observations, most
models invoke massive outflows driven by the radiation pressure generated by
quasars, which are expected to clear their host
galaxies of the bulk of their gas, therefore quenching star formation
\citep{Silk1998,Granato2004,DiMatteo2005,Springel2005,Croton2006,Narayanan2008,
Hopkins2008,Fabian2009,King2010,King2011,Hopkins2010,Zubovas2012}.
Evidence for such negative feedback has been found only recently in quasar
hosts through the discovery of massive outflows, in particular through
the detection of P-Cygni profiles of far-IR OH transitions
\citep{Fischer2010,Sturm2011}, the detection
high velocity wings of molecular emission
lines \citep[CO, HCN, ][]{Feruglio2010,Alatalo2011,
Aalto2012} and the detection of high velocity neutral gas in absorption
and of high velocity ionized gas \citep{Nesvadba2010,Rupke2011,Greene2011}.
Evidence for quasar-driven massive outflows have also been found through
the bright [CII]158$\mu$m fine
structure line; indeed,  Herschel spectra of nearby quasars
have revealed prominent [CII] broad wings,
associated with the molecular outflow traced by the OH
absorption lines
\citep[][Sturm et al. in prep.]{Fischer2010}. Quasar-driven
outflows have been detected also at high redshift
\citep[although exploiting different tracers,
][]{Alexander2010,Nesvadba2011,Harrison2012}.
Moreover, evidence has been found that these outflows are
indeed quenching star formation in their host galaxies
\citep{Cano-Diaz2012,Farrah2012,Steinhardt2011,Page2012},
as expected by models.

However, the discovery of old (age $>$ 2 Gyr), passive and massive galaxies at
z$\sim$2 \citep{Cimatti2004,Saracco2005,McCarthy2004,Glazebrook2004},
when the Universe had an age of only $\sim$3 Gyr,
requires that the quasar feedback quenching mechanism must
have been at work already at z$>$6 (age of the Universe less than one Gyr),
i.e. close to the re-ionization epoch.
Quasar driven winds have been observed up to z$\sim$6
\citep{Maiolino2001,Maiolino2004a},
however these are associated with ionized gas in the vicinity of the black
hole, accounting only for a tiny fraction of the total gas in the host galaxy.
So far,
no observational evidence was found for the massive, quasar driven
outflows at z$>$6 required by feedback models to explain the population of old
massive galaxies at z$\sim$2.

Here we focus on one of the most distant
quasars known, SDSSJ114816.64+525150.3 (hereafter J1148+5251), at z=6.4189
\citep{Fan2003}.
CO observations
have revealed a large reservoir of molecular gas,
$\rm M_{H_2}\sim 2\times 10^{10}~M_{\odot}$,
in the quasar host galaxy \citep{Walter2003,Bertoldi2003b}. The
strong far-IR thermal emission inferred from
(sub-)millimeter observations reveals
vigorous star formation in the host galaxy, with $\rm SFR \sim 3000~
M_{\odot}~yr^{-1}$ \citep{Bertoldi2003a,Beelen2006}.
J1148+5251 is also the first high redshift galaxy in which the
the [CII]158$\mu$m line was discovered \citep{Maiolino2005}.
High resolution mapping of the same line with the IRAM PdBI
revealed that most of the emission is confined within $\rm \sim 1.5~kpc$,
indicating that most of the star formation is occurring within a very
compact region \citep{Walter2009}.

Previous [CII] observations of J1148+5251 did not have a bandwidth large enough
to allow the investigation of broad wings tracing outflows, as in local
quasars. In this letter we present new IRAM PdBI observations
of J1148+5251 that, thanks to the wide bandwidth offered by the new correlator,
have allowed us to discover broad [CII] wings tracing a very massive
and energetic outflow in the host galaxy of this early quasar.
We show that the properties of this outflow
are consistent with the expectations of quasar feedback models.

We assume the concordance $\Lambda$-cosmology with
$\rm H_0 = 70.3~km~s^{-1}~Mpc^{-1}$, $\rm \Omega_\Lambda = 0.73$
and $\rm \Omega _m = 0.27$ \citep{Komatsu2011}.

\section{Observations and data reduction}

Observations with the IRAM PdBI were obtained mostly in April 2011
in D configuration (mostly with $\rm 1.5<PWV<3.5~mm$),
while a few hours where also obtained in January 2011
in C+D configuration ($\rm PWV<1.5~mm$). The resulting synthesized beam is
$\rm 2.2''\times 1.8''$. The receivers were tuned to 256.172~GHz, which
is the rest frame frequency of [CII] at the redshift of the quasar,
z=6.4189 \citep{Maiolino2005}.
The following flux calibrators were used:
3C454.3, MWC349,  0923+392, 1150+497, 3C273, 3C345, 1144+542,
J1208+546, J1041+525,  1055+018. Uncertainties on the absolute
flux calibration are 20\%.
The total on-source integration time was 17.5~hours,
resulting in a sensitivity of 0.08~Jy~km~s$^{-1}$~beam$^{-1}$ in
a channel with width 100~km~s$^{-1}$.

The data were reduced by using the CLIC
and MAPPING packages, within the GILDAS-IRAM software.
Cleaning of the resulting maps was run by selecting the clean components on an area of
$\sim 3"$ around the peak of the emission. For each map the  resulting residuals
are below the 1$\sigma$ error, ensuring that sidelobes are properly cleaned away.
Anyhow, as discussed in the following, the size determination has been investigated
directly on the {\it uv} data, therefore independently of the cleaning.

\begin{figure}
a)\\{\centerline{\includegraphics[width=7.truecm]{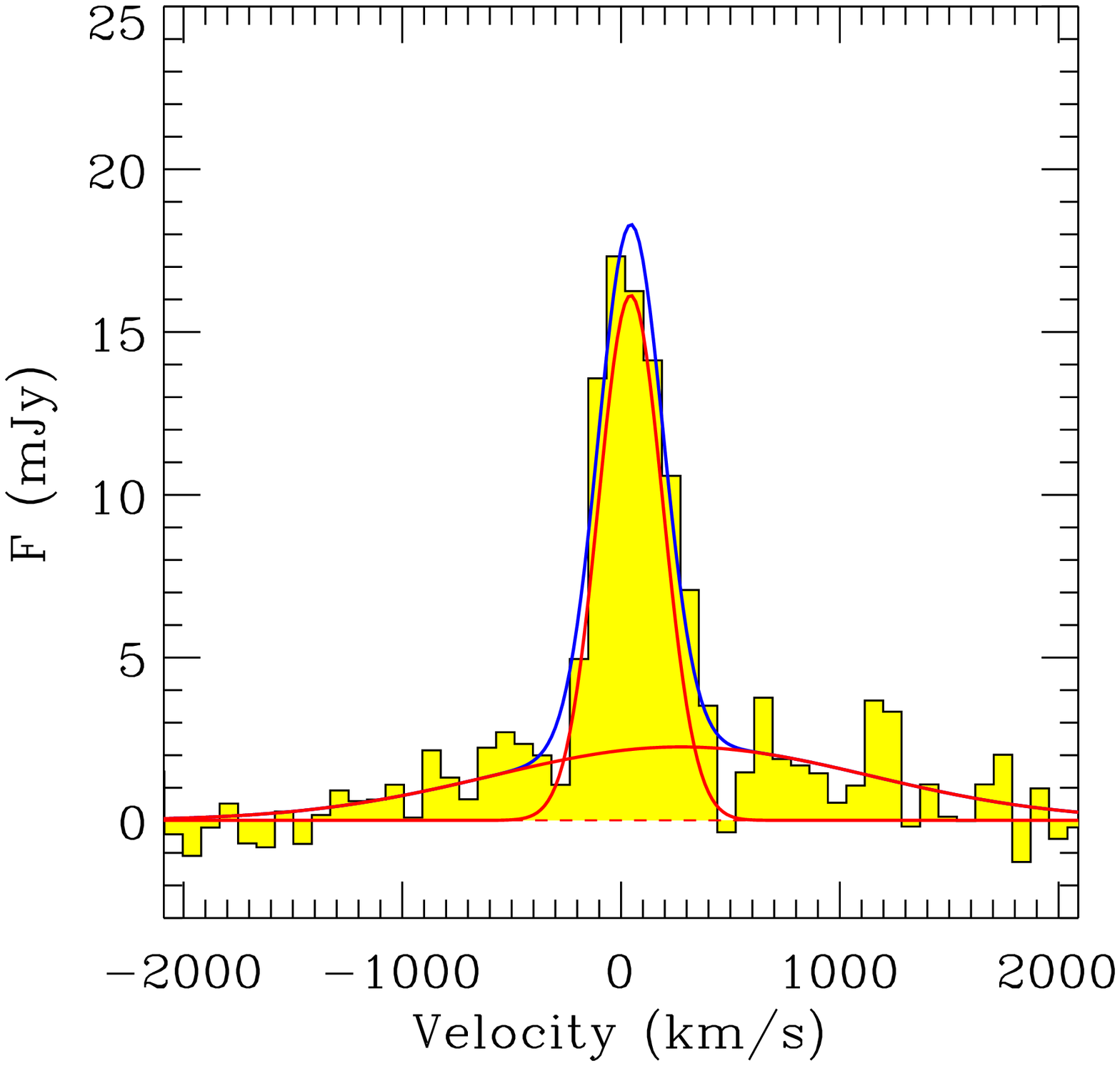}}}\\
b)\\{\centerline{\includegraphics[width=7.truecm]{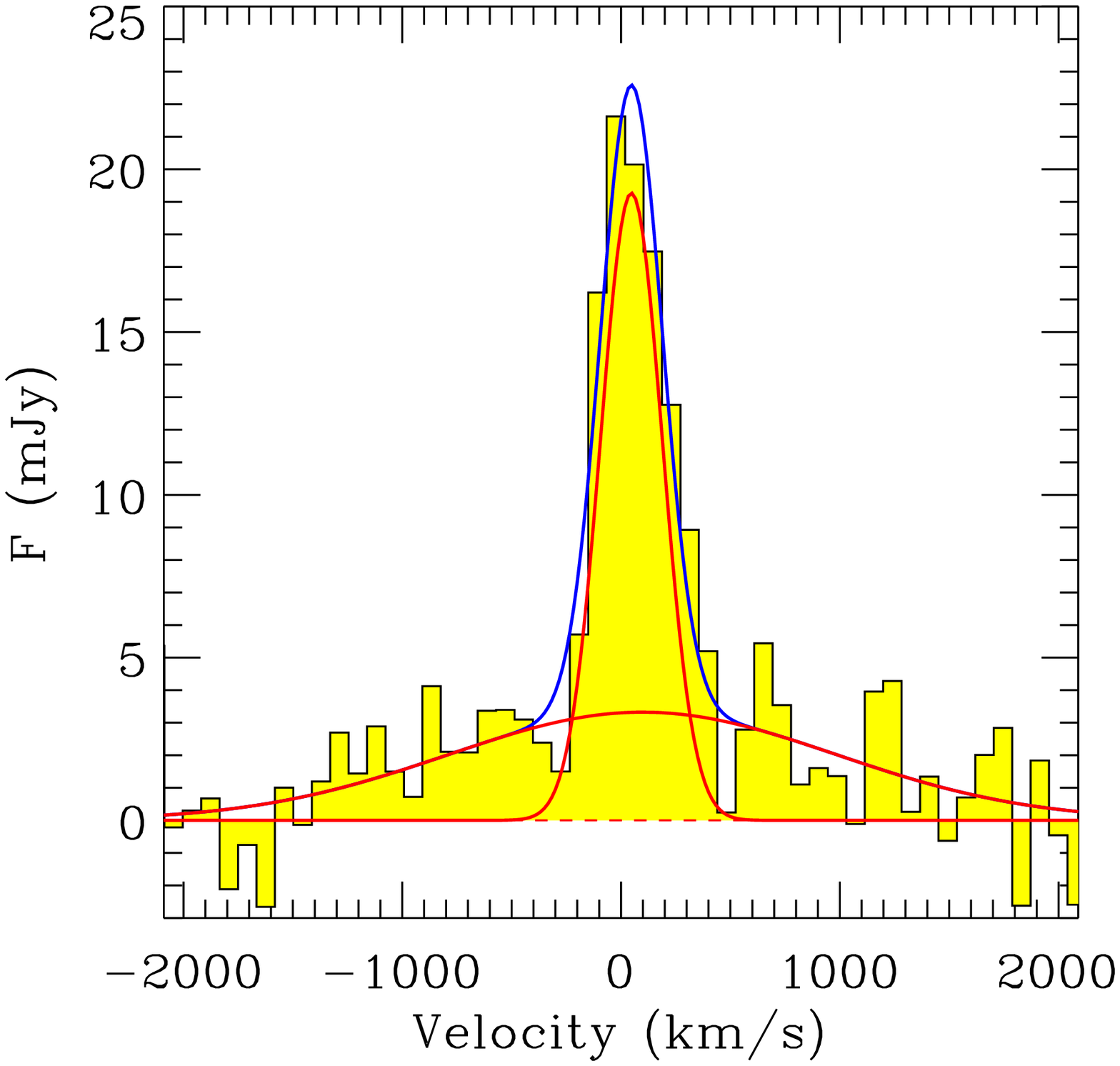}}}
%\subfloat[][]{\includegraphics[width=7.5truecm]{cii_spec4.eps}}\\
%\subfloat[][]{\includegraphics[width=7.5truecm]{cii_spec6.eps}}
\caption{IRAM PdBI continuum-subtracted spectrum of the [CII]158$\mu$m line,
redshifted to 256.172 GHz, in the host galaxy of the quasar J1148+5152
extracted from an aperture with a diameter
of 4$''$, top,  and 6$''$, bottom.
The spectrum has been resampled to a bin size of 85~km~s$^{-1}$.
The red lines show a double Gaussian fit
(FWHM=345~km~s$^{-1}$ and FWHM=2030~km~s$^{-1}$)
to the line profile, while the blue line shows the sum
of the two Gaussian components.}
\label{fig_spec}
\end{figure}

\section{Results}

\subsection{Detection of broad wings}

The continuum was subtracted from the {\it uv} data by estimating its level
from the channels at $\rm v<-1300~km~s^{-1}$ and at
$\rm v>+1300~km~s^{-1}$. The inferred continuum flux is 3.7 mJy,
which is fully consistent with
the value expected (4 mJy) from the bolometric observations \citep{Bertoldi2003a},
once the frequency range
of the latter and the steep shape of the thermal spectrum are taken into account.

Fig.~\ref{fig_spec}a shows
the continuum-subtracted spectrum, extracted from an aperture of 4$''$
(corresponding to a physical size of 11 kpc).
Fig.~\ref{fig_spec}b shows the spectrum extracted from a larger aperture of 6$''$ that,
although noisier than the former spectrum, recovers residual flux associated with the
beam wings and with any extended component.

The spectrum shows a clear narrow [CII]158$\mu$m emission line,
which was already detected by
previous observations \citep{Maiolino2005,Walter2009}.
However, thanks to the much wider bandwidth relative to
previous data, our new spectrum reveals broad [CII] wings extending
to about $\rm \pm 1300~km~s^{-1}$. These
are indicative of a powerful outflow, in analogy with the broad wings that have
been observed in the molecular and fine structure lines of local quasar host galaxies.
The spectrum
is fit\footnote{Note that, although Fig.~\ref{fig_spec} shows the spectrum resampled in channels of
$\rm 85~km~s^{-1}$ for sake of clarity, the fit was performed on the unbinned spectrum
to avoid any fitting artifact associated with the binning.}
with a narrow (FWHM=345 km s$^{-1}$)
and a broad (FWHM=2030 km s$^{-1}$) Gaussian, as shown in Fig.~\ref{fig_spec}, resulting
into  $\rm \chi^2_{red}=1.11$. By removing the broad component the
$\rm \chi^2$ increases from 195 to 229 (with 175 degrees of freedom),
implying that the broad component is required at a confidence level
higher than 99.9\%.

Theoretical models predict that starburst-driven winds cannot reach velocities
higher than 600 km s$^{-1}$ \citep{Martin2005,Thacker2006},
therefore the high velocities observed in the
[CII] wings of J1148+5251 strongly favor radiation pressure from the quasar
nucleus as the main driving mechanism of the outflow. Quasar radiation
pressure is favored, relative to SN-driven winds, also based on
energetics arguments, as discussed later on.

\subsection{Outflowing gas mass}

The luminosity of the broad [CII] component allows us to infer a lower limit
of the outflowing atomic gas mass, using,
\begin{equation}
\rm
\frac{M_{outfl}(atomic)}{M_{\odot}} = 0.77~
\left(\frac{0.7~L_{[CII]}}{L_{\odot}}\right)
\left(\frac{1.4~10^{-4}}{X_{C^+}}\right)
\frac{1+2~e^{-91K/T}+n_{crit}/n}{2~e^{-91K/T}}
\label{eqCII}
\end{equation}
\citep{Hailey2010}, where $\rm X_{C^+}$ is the $\rm C^+$ abundance per
hydrogen atom, $\rm T$ is the gas temperature, $\rm n$ is the gas density
and $\rm n_{crit}$ is the critical density of the [CII]158$\mu$m
transition ($\rm \sim 3~\times ~10^3~cm^{-3}$). By assuming a density much higher
than the critical density, Eq.~\ref{eqCII} gives a lower limit on the
mass of atomic gas. The quasar-driven outflows observed locally are
characterized by a wide range of densities, including both dense clumps
and diffuse, low density
gas \citep{Cicone2012,Aalto2012,Fischer2010},
therefore our assumption on the gas density gives a conservative lower limit
on the outflowing gas mass. We assume a temperature of 200~K, however
the dependence on the temperature is weak (going from 100~K to
1000~K the implied gas mass is within 20\% of the value obtained
at 200~K). Furthermore, we assume a $\rm C^+$ abundance typical of PDRs,
i.e. $\rm X_{C^+}=1.4~\times ~10^{-4}$ \citep{Savage1996}, which is also conservative,
since the gas in the outflow is certainly, on average, at a higher
state of ionization.
The luminosity of the [CII] broad component is inferred from
the flux of this component in the spectrum extracted from an aperture
of 6$''$ ($\rm F_{[CII]}(broad)=6.8~Jy~km~s^{-1}$), yielding
$\rm L_{[CII]}(broad) = 7.3\pm 0.9~\times 10^9~L_{\odot}$.
We obtain a lower limit on
the outflowing atomic gas mass of
\begin{equation}
\rm M_{outfl}(atomic) > 7 ~\times ~10^9 ~M_{\odot}
\label{eqMout}
\end{equation}
We emphasize that this is a conservative lower limit on the total mass of
outflowing gas, not only because of the assumptions on the physical properties
of the outflowing atomic gas, but also because a significant fraction of the
outflowing gas is likely in the molecular form.

\begin{figure}
\includegraphics[width=4.2truecm]{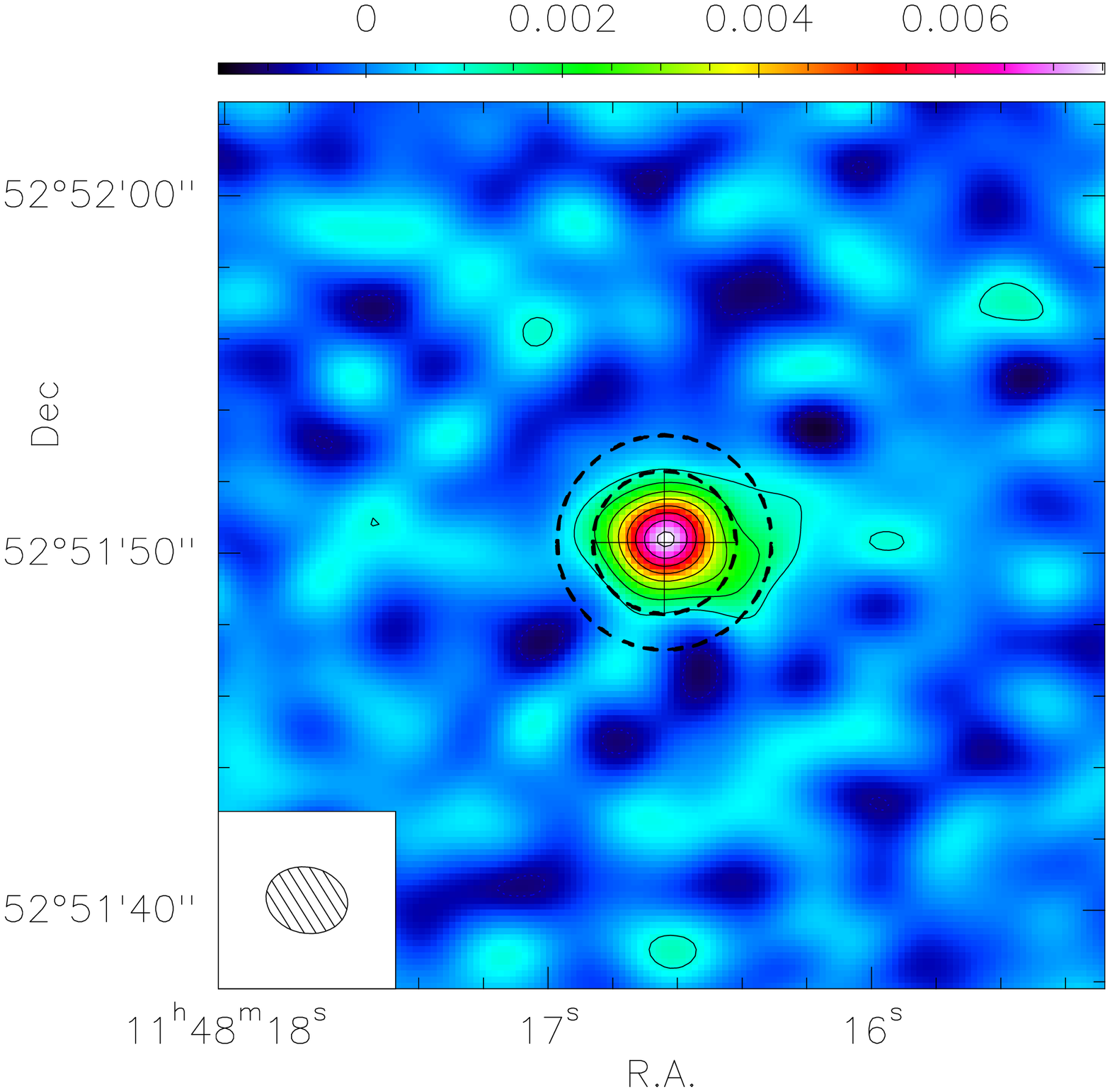}
\includegraphics[width=4.2truecm]{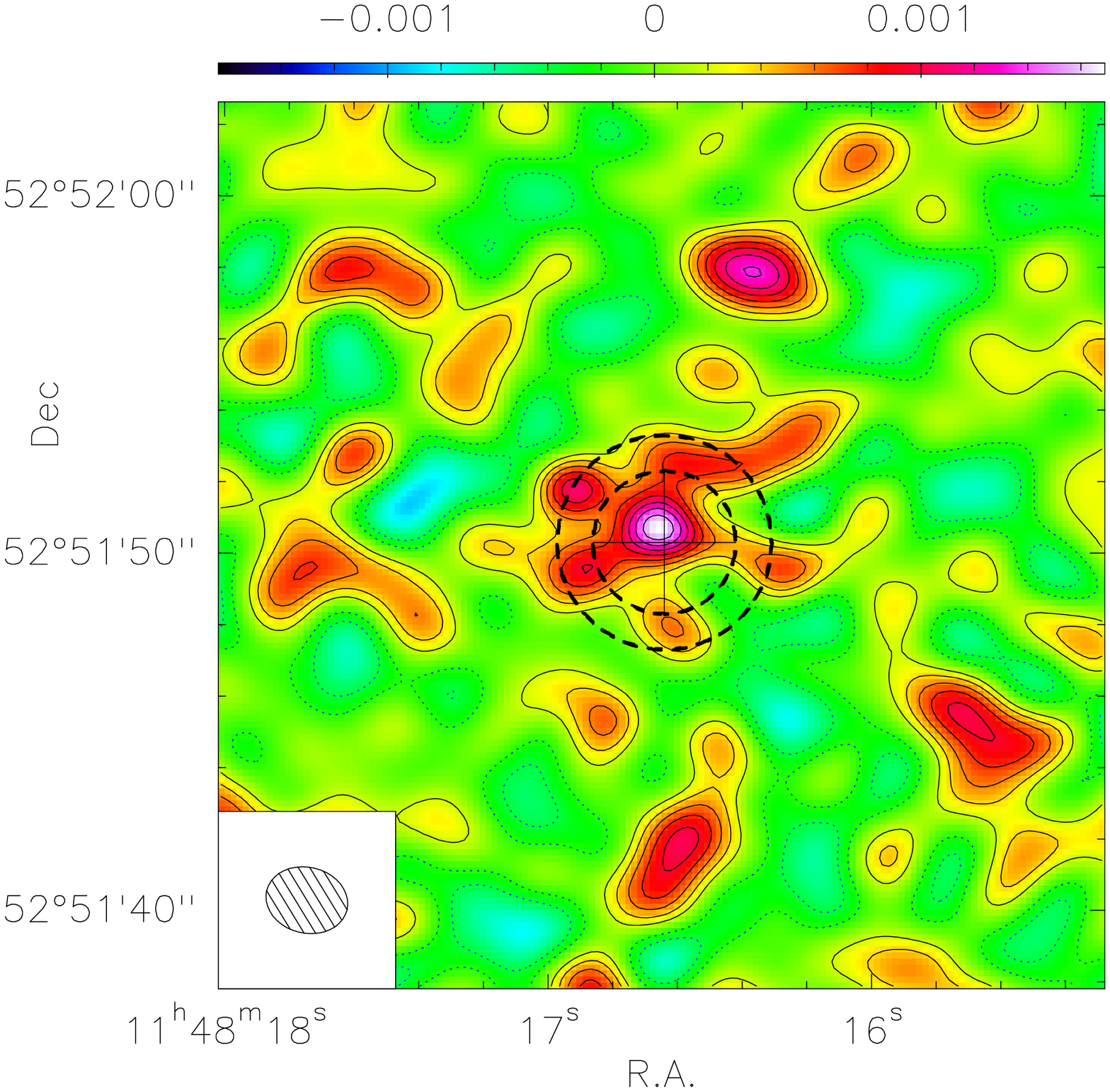}
\caption{
Map of the continuum-subtracted [CII] line narrow component (a),
$\rm -300<v<+400~km~s^{-1}$, and of the [CII] line wings (b),
$\rm -1300<v<-300~km~s^{-1}$ and $\rm +400<v<+1300~km~s^{-1}$. See text for
details on the continuum subtraction of the narrow component.
The dashed circles indicate the extraction apertures of the two spectra
shown in Fig.~\ref{fig_spec}.
Levels are in steps of  $\rm 0.64~Jy~km~s^{-1}~beam^{-1}$ (i.e. 3$\sigma$)
in the narrow component map (a) and in steps of $\rm 0.36~Jy~km~s^{-1}~beam^{-1}$
(i.e. 1$\sigma$) in the broad wings map (b).
The beam of the observation is shown in the bottom-left corner.
At the redshift of the source 1$''$ corresponds to 5.5~kpc. The colorbars are
in units of Jy beam$^{-1}$, giving the average flux density in each velocity
integration interval;
to obtain fluxes in units of Jy km s$^{-1}$ beam$^{-1}$ one has to
multiply by the width of the integration interval of each map.
}
\label{fig_map}
\end{figure}

\subsection{Extension}

Fig.~\ref{fig_map}(a) shows the map of the [CII] narrow component
integrated within\footnote{The line core integration
limits are asymmetric because the narrow component is slightly
skewed towards positive velocities.}
$\rm -300<v<+400~km~s^{-1}$.
Note that in this case we have subtracted a pseudo-continuum
defined by the flux observed at $\rm 400<|v|<800~km~s^{-1}$ (resulting
in a level of 5.6~mJy),
to minimize the contribution of the broad component.
Fig.~\ref{fig_map}(b) shows the map of the [CII] broad wings,
where we have combined the blue
$\rm (-1300 < v < -300~km~s^{-1})$ and red
$\rm (+400 < v < +1300~km~s^{-1})$
wings to improve the signal-to-noise.

\begin{figure}
\includegraphics[width=3.8truecm,angle=-90]{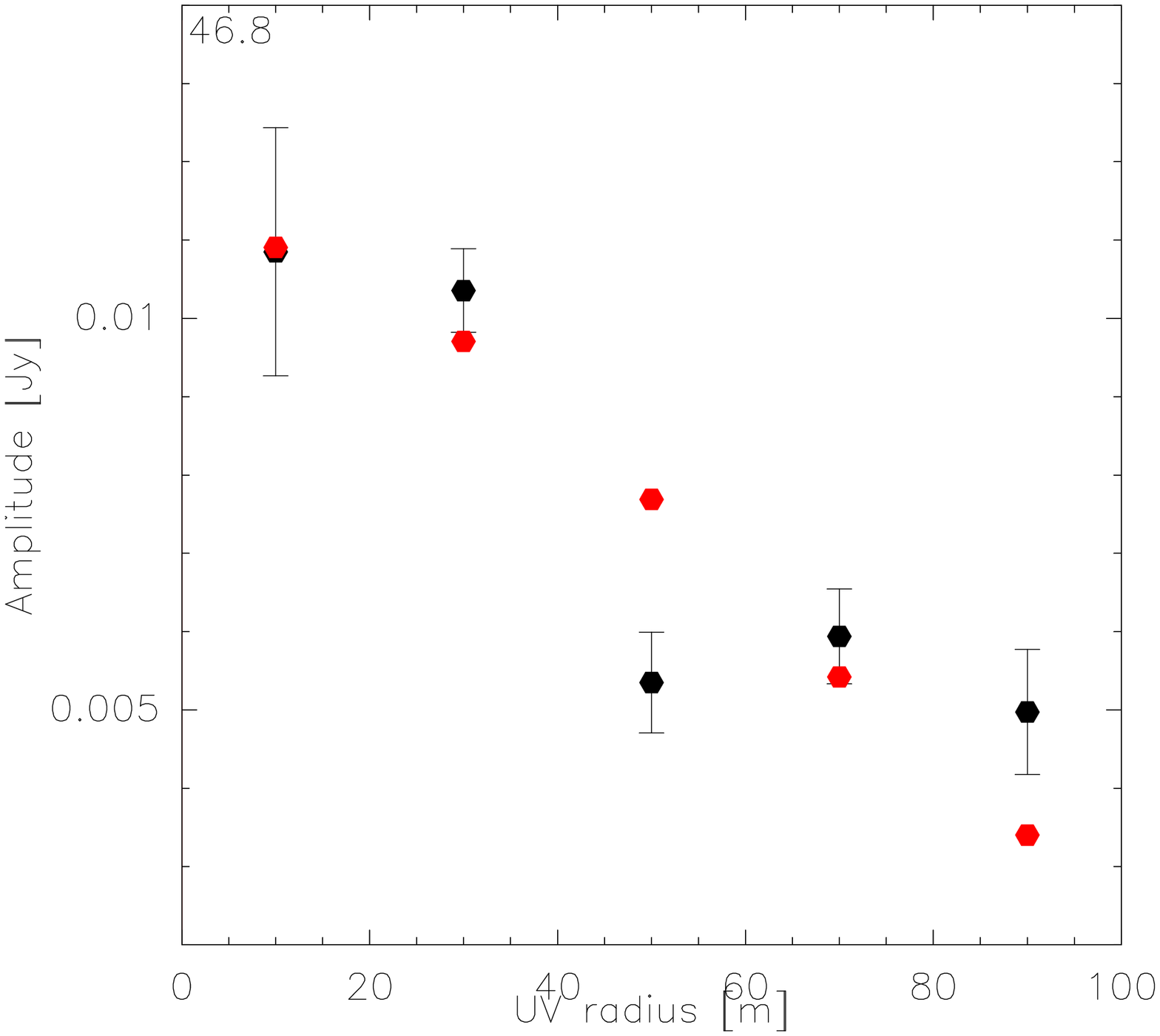}
\includegraphics[width=3.8truecm,angle=-90]{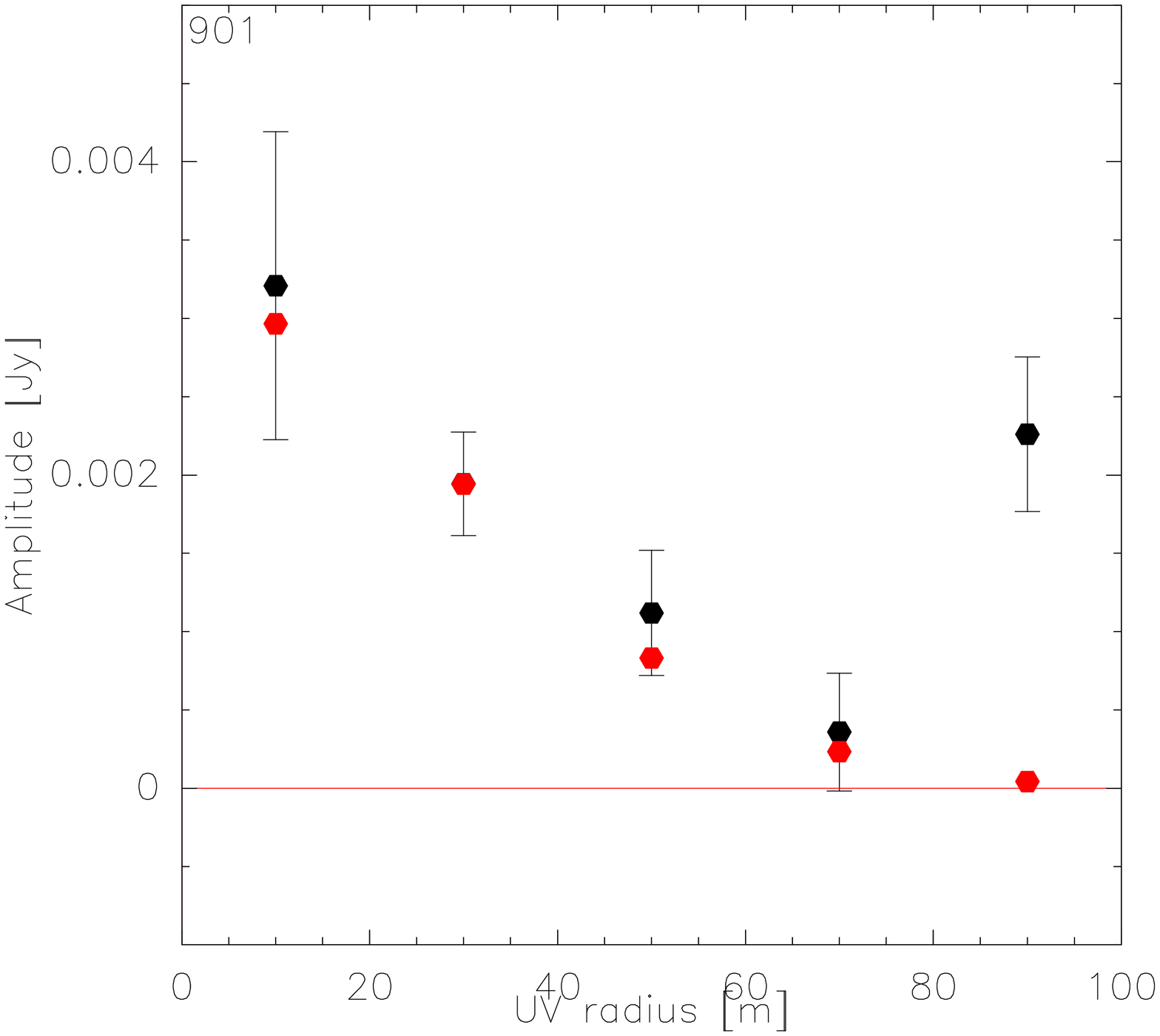}
\caption{
Amplitude of the visibilities as a function of the baseline distance (black symbols) for the
narrow component of the [CII] line (a) and for its broad wings (b).
In both cases the distribution of visibilities is inconsistent with an unresolved
source (which would give a flat distribution).
The reds symbols show the result of the best fit by using a circular Gaussian model.
Note that the units (Jy) refer to the average flux in the selected channels associated with
the line, to obtain the units in terms of line flux one has to multiply by 700~km~s$^{-1}$
and by 1800~km~s$^{-1}$, respectively.
}
\label{fig_uv}
\end{figure}

The narrow component
(Fig.\ref{fig_map}a), tracing gas and star formation in the host galaxy is resolved at high
significance.
To illustrate this extension more quantitatively, Fig.~\ref{fig_uv}a shows
the amplitude of the visibilities as a function of the baseline
distance (black symbols) for the narrow component of the [CII] line.
Errorbars show the statistical photon noise on the average amplitude in each bin.
In this diagram an unresolved source has a flat distribution. Such a flat distribuion can
be ruled out by the {\it uv} data.
A fit of the {\it uv} data with a simple circular Gaussian
gives a size FWHM=1.5$''$, corresponding to 8~kpc.
However, as illustrated by the
red symbols, showing the result of the fit, a single circular Gaussian does not
provide a good description of the data, indicating that the morphology of
the [CII] core emission is more complex.
We know that the [CII] core has a very compact component (size 0.3$''$, \citealt{Walter2009})
which is unresolved at our
resolution. This compact component is responsible for the flat visibilities at
baselines longer than 40 m. The rising visibilities at baselines shorter than 40
m indicate the presence of a resolved component, with size larger than 2$''$, whose
flux is about 40\% of the total narrow line flux. It is likely that this extended component of the
narrow [CII] emission was resolved out in previous high angular resolution
observations.

The spatial extension of the line wings is more difficult to quantify due the lower
signal-to-noise. The map
of the wings (Fig.\ref{fig_map}b)
is characterized by a compact core, detected at 8$\sigma$, and diffuse emission surrounding
the source on scales of a few arcsec, but at low significance. The amplitude versus {\it uv}-radius
diagram, shown in Fig.~\ref{fig_uv}b, suggests that the the outflow is indeed marginally resolved, as indicated
by the decreasing visibilities out to 70~m. The visibilities at distances $<$70m are inconsistent
with a flat distribution at confidence level higher than 99\%. A fit of the
{\it uv} visibilities with a simple circular Gaussian (shown with the red symbols)
gives a size of 2.9$''$ (16 kpc). If confirmed (with higher resolution and deeper observations)
this would be the largest outflow ever detected and it would support models ascribing the capability
of cleaning entire galaxies, out to large radii, to quasar-driven outflows.
However, the intensity of the last visibility at 90~m
suggests that the structure of the outflow is more complex and
cannot be modelled with a simple Gaussian. We have attempted to fit a ring-like morphology to the
uv data, but the fit does not improve and the inner radius is set to a very small value ($<$0.1$''$).

Finally, we note that the fit also requires an offset of the core of the wings of 0.3$''$
($\pm$0.1$''$)
relative to the line core, this is also observed in the map (Fig.~\ref{fig_map}b).
The offset is small
and marginally significant. However, even if real, an offset of the centroid of the outflowing gas
is not surprising, since outflows observed in local and high-z galaxies are not aligned with the
galaxy centers \citep{Weis2001,Walter2004b,Wilson2008,Alatalo2011}.

\subsection{Outflow rate, energetics and depletion time}

Given the limited information available, we cannot elaborate on sophisticated models. We have assumed a
simple scenario where the outflow occurs in a spherical volume
(or conical or multi-conical volume), with radius
$\rm R_{outflow}$,
uniformly filled with outflowing clouds. Note that the outflowing
clouds can have any density distribution,
since we are assuming the conservative case of $\rm n\gg n_{cr}$, which gives a lower
limit on the outflowing mass according to Eq.1. The presence of clouds with density below
the critical density would result into an even higher outflow mass and outflow rate.

For the geometry discussed above, the volume-averaged density of atomic
gas in the outflow is
$
\rm \langle \rho _{outfl} \rangle _V = 3~M_{outfl}/(\Omega R^3_{outfl})
$
where $\Omega$ is the total solid angle subtended by the (multi-)conical outflow
($\rm \Omega = 4\pi$ for a spherical outflow). Note that
$\rm \langle \rho _{outfl} \rangle _V$ should not be confused with the gas
density of the individual clouds.
In the case of a shell
(or fragmented shell) geometry the volume-averaged density is
$
\rm \langle \rho _{outfl} \rangle _V = 3~M_{outfl}/[\Omega (R^3_{outfl}-R^3_{in})]
\approx M_{outfl}/(\Omega~R^2_{outfl}~ dR)
$,
where $\rm R_{in}$ is the inner radius of the shell and where the last approximation
holds in the case that the shell thickness ($\rm dR=R_{outfl}-R_{in}$)
is much smaller than its radius.
In these equations $\Omega$ is the total solid angle subtended by the
(fragmented) shell.

In the case of a spherical, or multi-conical, geometry
the outflow rate is then given by
\begin{equation}
\rm \dot{M}_{outfl} \approx {\mathit v}~\Omega ~R^2_{outfl}~\langle \rho _{outfl}\rangle _V
	= 3~{\mathnormal v}~\frac{M_{outfl}}{R_{outfl}}
\end{equation}
where $\mathnormal v$ is the outflow velocity.
In the case of a (fragmented) shell-like geometry
\begin{equation}
\rm \dot{M}_{outfl} \approx 3 {\mathnormal v}~ \frac{R^2_{outfl}~M_{outfl}}{R^3_{outfl}-R^3_{inner}}
	\approx {\mathnormal v}~\frac{M_{outfl}}{dR} ~.
\end{equation}
Clearly the shell-like geometry gives an even higher outflow rate relative to the
spherical/multi-conical geometry. In the following we conservatively assume the latter geometry.
Note also that the outflow rate is independent of $\Omega$, both in the
(multi-) conical and in the (fragmented) shell geometries.

Since we have a lower limit on $\rm M_{outfl}$,
we can derive a conservative lower limit on the outflow rate.
In particular, by taking $\rm R_{outfl}$ = 8 kpc and
$\rm v =1300~km~s^{-1}$ (conservatively assuming that the maximum velocity observed in the
wings is the de-projected velocity of the outflow, which is a lower limit in the case of
a conical outflow not intercepting the plane of the sky), we obtain
\begin{equation}
\rm \dot M_{outfl} > 3500~M_{\odot}~yr^{-1}.
\label{eq_llMout}
\end{equation}
%This is the highest outflow rate ever found.
We note that
if the outflow size has been overestimated, due to the low signal-to-noise in the wings, as discussed
above, and the bulk of the flux is actually unresolved, then the lower limit on the outflow
rate in Eq.~\ref{eq_llMout} is even more conservative, since the outflow rate scales as $\rm 1/R_{outfl}$.
As discussed above, a (fragmented)
shell-like geometry would given a even higher outflow rate.
Concerning the velocity field, the angular resolution of our data does not allow us to map the
kinematics of the outflow. However, we note that, in the 
nearby quasar Mrk231, transitions tracing the outflow
on different scales have the same profile and the same maximum velocity
\citep{Aalto2012,Cicone2012},
supporting the scenario where quasar driven outflows have velocity roughly
constant throughout the wind. The
same kinematics is likely to apply to J1148 as well. However, new higher angular resolution observations
will allow us to better constrain the kinematics.

The inferred
lower limit on the outflow rate is similar to the star formation rate in the host galaxy
($\rm \sim 3000~M_{\odot}~yr^{-1}$).

The kinetic power associated with the outflow is given by
\begin{equation}
\rm P_K \approx 0.5 ~v^2~ \dot{M}_{outfl} > 1.9~\times ~10^{45}~erg~s^{-1}
\end{equation}
which is about 0.6\% of the bolometric luminosity of the quasar nucleus.
On the one hand, this is well below the maximum efficiency expected for quasars to drive
outflows ($\rm P_K/L_{bol}\sim 5\%$, \citealt{Lapi2005}),
meaning that the quasar radiation pressure is probably fully capable of driving
the outflow observed in J1148+5251. On the other hand, the lower limit on the
outflow kinetic power is barely consistent with the kinetic
power that can be achieved by a starburst driven wind
($\rm P_K(SB) = 7~\times ~10^{41}~(SFR/M_{\odot}~yr^{-1})~erg~s^{-1} \approx 2 ~\times ~10^{45}~erg~s^{-1}$
for J1148+5152,
\citealt{Veilleux2005}), therefore we cannot exclude a starburst
contribution to the wind based solely on energetic arguments and by
using simple scaling equations. However,
as discussed above the very high outflow velocities are difficult to explain
by starburst driven winds models, hence strongly favoring quasar radiation
pressure as the main driving mechanism.

We also mention that a parallel
paper \citep{Valiante2012}, by exploiting
existing detailed cosmological models specifically
suited for J1148 constrained by all observable quantities,
do predict a quasar-driven wind fully consistent with the outflow rate
measured by us, while the outflow contribution by the starburst is orders
of magnitude lower.

The molecular gas content in the host galaxy of J1148+5251, as inferred by CO observations,
is $\rm 2~\times ~10^{10}~ M_{\odot}$ \citep{Walter2003,Bertoldi2003b}.
At the observed (minimum) outflow rate, the quasar host galaxy will be cleaned of its gas content,
hence quenching star formation, within less than $\rm 6~\times ~10^6~yr$.

\section{Conclusions}

By using the IRAM PdBI,
we have discovered an exceptional outflow in the host galaxy of the quasar J1148+5251
at z=6.4, through the detection of prominent broad wings of the [CII]158$\mu$m line,
with velocities up to about $\rm \pm 1300~km~s^{-1}$.
This is the most distant massive outflow ever detected and it is
likely tracing the long sought quasar-feedback already in
place in the early Universe, close to the re-ionization epoch.
Both the high outflow velocity and the
kinetic power ($\rm P_K > 1.9~\times ~10^{45}~erg~s^{-1}$)
favor quasar radiation pressure as the main driving mechanism.
A more detailed cosmological model of J1148 \citep{Valiante2012}
confirms that
in this object a quasar-driven outflow is expected with an outflow
rate consitent with what observed by us,
while the contribution from the starburst wind
is orders of magnitude lower.

We marginally resolve the outflow, with a size of about 16~kpc, which would make it the largest
outflow ever detected. This finding supports models ascribing the capability of cleaning
entire galaxies, out to large radii, to quasar-driven outflows. However, higher angular
resolution and deeper observations are required to confirm the outflow extension.

We infer a conservative lower limit on the total outflow rate of
$\rm \dot{M}_{outfl}>3500~M_{\odot}~yr^{-1}$, which would be the highest outflow rate
ever detected. This lower limit on the outflow rate is higher than the SFR in the
quasar host ($\rm SFR\approx 3000~M_{\odot}~yr^{-1}$). At the observed outflow rate
the gas content in the host galaxy will be cleaned, and therefore star formation
will be quenched, in less than $\rm 6\times 10^6~yrs$.
Such a fast and efficient quenching mechanism, already at work at z$>$6,
is what is required by models to explain the properties
of massive, old and passive galaxies
observed in the local Universe and at $\rm z\sim 2$.

\section*{Acknowledgments}

This work is based on observations carried out with the IRAM Plateau de Bure
Interferometer. IRAM is supported by INSU/CNRS (France), MPG (Germany) and IGN
(Spain). We are grateful to A. Marconi for useful comments. S.G. thanks INAF
for support through an International Post-Doctoral Fellowship.
%C.C. thanks the University of Cambridge for support.

%\bibliographystyle{mn2e.bst}
\setlength{\labelwidth}{0pt}
\bibliographystyle{mn2e}
\bibliography{bibl3}
%\begin{thebibliography}{99}
%\end{thebibliography}

%\bsp

%\label{lastpage}

\end{document}